\documentclass[twocolumn,nofootinbib]{revtex4}%

\usepackage{amssymb}
\usepackage{amsfonts}
\usepackage{amsmath}
\usepackage{graphicx,color}
\usepackage{slashed}
\usepackage{young}
\usepackage[vcentermath]{youngtab}
\usepackage[utf8]{inputenc}

\usepackage{tensor}
\usepackage{hyperref}
\usepackage{ulem}

\makeatletter

\newcommand{\be}{\begin{equation}}
\newcommand{\ee}{\end{equation}}
\newcommand{\bea}{\begin{eqnarray}}
\newcommand{\eea}{\end{eqnarray}}

\renewcommand{\tilde}{\widetilde}
\renewcommand{\hat}{\widehat}

\renewcommand{\d}{\partial}

\@addtoreset{equation}{section}

\newcommand*\xbar[1]{%
  \hbox{%
    \vbox{%
      \hrule height 0.5pt 
      \kern0.3ex
      \hbox{%
        \kern-0.0em
        \ensuremath{#1}%
        \kern-0.0em
      }%
    }%
  }%
} 

\begin{document}

\title{Asymptotic realization of the super-BMS algebra at spatial infinity}
\author{Marc Henneaux${}^{1,2}$, Javier Matulich${}^1$ and  Turmoli Neogi${}^1$}
\affiliation{${}^1$Universit\'e Libre de Bruxelles and International Solvay Institutes, ULB-Campus Plaine CP231, B-1050 Brussels, Belgium}
\affiliation{${}^2$Coll\`ege de France, 11 place Marcelin Berthelot, 75005 Paris, France}

\email{henneaux@ulb.ac.be, jmatulic@ulb.ac.be, turmoli.neogi@ulb.ac.be}

\begin{abstract}
{Explicit boundary conditions are given at spatial infinity for four-dimensional supergravity, which provide a realization of the super-BMS algebra of Awada, Gibbons and Shaw.  The results are then generalized to the $N$- extended super-BMS algebras.}
\end{abstract}



\maketitle

\section{Introduction}
\setcounter{equation}{0}

The super-BMS algebra was introduced in \cite{Awada:1985by} and shown there to be the asymptotic symmetry algebra of supergravity with zero cosmological constant. It is a graded extension of the BMS algebra \cite{Bondi:1962px,Sachs:1962wk,Sachs:1962zza,Penrose:1962ij,Madler:2016xju,Alessio:2017lps,Ashtekar:2018lor} with the striking feature that it contains only four real fermionic generators.  Out of the infinite number of BMS supertranslations, only the ordinary, angle-independent, translations have a fermionic square root.

The analysis of \cite{Awada:1985by} was carried out at null infinity. This seemed natural at the time since it was thought then that the BMS symmetry was inseparable from gravitational radiation and could therefore only be exhibited at $\mathcal{I}^+$ or $\mathcal{I}^-$. However, the BMS symmetry is a true symmetry of the theory, with bona fide conserved Noether charges, and these can be written down in any formulation.   In particular, the BMS symmetry can be identified at spatial infinity \cite{Henneaux:2018cst,Henneaux:2018hdj,Henneaux:2019yax}, a fact that has a direct implication on the description of the physical states, generally defined on Cauchy hypersurfaces.  Displaying the BMS algebra at spatial infinity is also quite important as it disantagles the BMS symmetry from the intricate dynamical question of the existence of a null infinity with the necessary smoothness properties \cite{Christodoulou:1993uv,Friedrich:2017cjg}. As a side historical remark, we note that the earlier foundational work \cite{Regge:1974zd,Ashtekar:1978zz,Beig1982} made crucial progress in the understanding of the asymptotic structure of gravity at spatial infinity.  With the boundary conditions considered in these insightful papers,  the symmetry group was found to be either smaller than the BMS group and just the Poincar\'e group, or bigger and infinite-dimensional (``Spi group'' of  \cite{Ashtekar:1978zz}) but with no corresponding standard moment map (but see \cite{Prabhu:2019daz}).  The reconciliation between null and spatial infinity results was achieved in \cite{Henneaux:2018cst,Henneaux:2018hdj,Henneaux:2019yax}.

Contrary to what one might have wrongly anticipated, the boundary conditions implementing the BMS symmetry at spatial infinity turn out to be conceptually and technically extremely simple. These are just the boundary conditions given in the pioneering work \cite{Regge:1974zd}, with a twist in the parity conditions involving an improper gauge transformation \cite{Benguria:1976in} written in Hamiltonian form.   This twist is what makes the BMS group act non trivially.  [We adopt for definiteness the boundary conditions of \cite{Henneaux:2018hdj,Henneaux:2019yax}.  Earlier investigations \cite{Henneaux:2018cst} analysed a different, more drastic twist in the parity conditions, inspired by \cite{Compere:2011ve}, and yielding also the BMS symmetry at spatial infinity. Although equally compatible with supersymmetry -- as it can easily be verified --,  these original boundary conditions present, however, some limitations (singular behaviour at null infinity, exclusion of the Taub-NUT solution) and have been replaced for that reason in  \cite{Henneaux:2018hdj,Henneaux:2019yax} by different ones which do not present these features. These alternative boundary conditions are characterized by an inequivalent and simpler twist in the parity conditions.  The article \cite{Henneaux:2019yax} reviews these more  recent boundary conditions on which the present work relies.]

The purpose of this note is to extend the analysis of \cite{Henneaux:2018hdj,Henneaux:2019yax} to cover supergravity.  In Section \ref{subsec:Starting}, we provide boundary conditions on the gravitino at spatial infinity which are consistent with the boundary conditions on the graviton.  We show then in Section \ref{sec:AsSymm} that these boundary conditions are invariant under the super-BMS algebra of \cite{Awada:1985by} and write the corresponding surface generators.  The Poisson bracket algebra of these generators is explicitly displayed in Section \ref{sec:AsymptoticAlgebra}. Section \ref{sec:ExtenstedSUSY} is devoted to extended supergravities. We finally comment in Section \ref{sec:Conclusions} on possible generalizations of the current work yielding a richer graded extension of the BMS algebra, with an infinite number of fermionic generators.  Appendix \ref{sec:appSusyBra} explicits some technical steps in the derivation of the Poisson bracket algebra of the supersymmetry generators.

As it is common practice in asymptotic investigations of the type carried here, we shall
assume ``uniform smoothness'' \cite{Sachs:1962zza} whenever needed, i.e., $\d_r o(r^{-k}) =
o(r^{-k-1}), \d_A o(r^{-k}) = o(r^{-k})$.  Similarly, the distinction between $O(r^{-(k+1)})$ and $o(r^{-k})$ will usually not be important to the orders relevant to the analysis.

\section{Action and boundary conditions}
\label{subsec:Starting}
\setcounter{equation}{0}

\subsection{Action}

We start with the action of $N=1$ supergravity in four spacetime dimensions, written in canonical form \cite{Teitelboim:1977fs,Tabensky:1977ic,Fradkin:1977wv,Deser:1977ur,Pilati:1977ht} (see also \cite{Henneaux:1978wlm})
\begin{eqnarray}
	&& S_H[e^a_i, \pi^i_a, \psi_k; N, N^k, \psi_0, \lambda_{ab}]  = \nonumber \\ &&  \int dt  \int d^{3}x \, 
		\left(\pi^i_a \d_t
		e^a_i  + \frac{i}{2}  \sqrt{g}  \psi_k^T \gamma^{km} \d_t \psi_m \right) \nonumber \\
		 	&&   \hspace{-0.5cm}   -\int dt  \int d^{3}x \left(   N \mathcal{H} + N^i \mathcal{H}_i + i \psi_0 ^T \mathcal{S} + \frac12 \lambda_{ab} \mathcal{J}^{ab} \right) \nonumber \\
&& \hspace{.5cm} + \int dt B_\infty \label{eq:StartingPoint}
\end{eqnarray}
where $B_\infty$ is a surface term at spatial infinity ($r \rightarrow \infty$),
which depends on the boundary conditions and which will be discussed below.  We impose the ``time gauge'' condition in which the timelike vector of the local frame (``vierbein'') is orthogonal to the hypersurfaces of constant time (our conventions are collected in  \cite{Conventions}).
The dynamical variables to be varied in the action principle are the spatial components $e^a_i$ of the triad (dreibein) ($a, b , \cdots =  1, 2, 3$ are internal $SO(3)$ indices, $i, j, \cdots = 1,2,3$ are spatial coordinate indices),  their conjugate momenta $\pi^i_a$, the spatial components $\psi_k$ of the gravitino field and the Lagrange multipliers $N$ (lapse), $N^k$ (shift), $\psi_0$ (temporal component of the gravitino field) and $\lambda_{ab}$ which implement the constraints
\be
\mathcal{H}  \approx 0, \quad \mathcal{H}_ i   \approx 0, \quad \mathcal{S}  \approx 0, \quad \mathcal{J}^{ab} \approx 0 .
\ee
We use as usual the symbol $\approx$ to denote equality on the constraint surface.   The constraints are all first class and are related to the gauge symmetries of the theory: diffeomorphisms, local supersymmetry, and internal $SO(3)$ rotations.

The constraints $\mathcal{H}  \approx 0$ and $ \mathcal{H}_ i \approx 0$ are the Hamiltonian and momentum constraints, and generate  diffeomorphisms.  They read
\be
\mathcal{H}  = \frac{1}{\sqrt{g}} \left( \pi_{ij} \pi^{ij} - \frac12 \pi^2 \right) - \sqrt{g} R + F_1
\ee
and
\be
\mathcal{H}_i  = - 2 {{\pi_i}^j}_{\vert j} + F_2
\ee
where 
\be
\pi^{ij} = \frac12  e^{a(i}\pi^{j)}_a  
\ee
Here, $F_1$ and $F_2$ are fermionic terms, which are at least bilinear in the fermions (the bilinears are typically of the form $\psi \d \psi$, or $\omega \psi \psi$ where $\omega$ is the spin connection, or $\pi \psi \psi$).   Their explicit expression can be found in \cite{Fradkin:1977wv,Deser:1977ur,Pilati:1977ht} and will not be needed here.

Similarly, the fermionic constraint $\mathcal{S}  \approx 0$ generates local supersymmetry and is
\be
\mathcal{S}  = \sqrt{g} \gamma^{mn} \d_m \psi_n + F_3 + F_4, 
\ee
where $F_3$ is linear in the fermions and in the spatial spin connection ($\sim \omega_{abm} \psi_k$) or the conjugate momentum $\pi^i_a$ ($\sim \pi^i_a \psi_m$) and where $F_4$ is at least cubic in the fermions.  Again we refer to \cite{Fradkin:1977wv,Deser:1977ur,Pilati:1977ht}  for the explicit expressions.

The constraints $\mathcal{J}^{ab} \approx 0$  generate local spatial rotations of the orthornormal frames (only $SO(3)$ transformations appear because we impose the time gauge condition $e^0_k = 0$, see  \cite{Conventions}).  They are algebraic in the fields (no derivative) and can be easily written down from the transformation rules of the fields under local rotations \cite{Fradkin:1977wv,Deser:1977ur,Pilati:1977ht}.  One has
\be
\delta e^a_i = {\omega^a}_b e^b_i, \qquad \delta \psi_k = \frac14 \omega_{ab} \gamma^{ab} \psi_k
\ee
for infinitesimal rotations $\omega_{ab} = - \omega_{ba}$. [If one were to convert the world index $k$ into a local index $c$, $\psi_c = e^k_c \psi_k$, one would find $\delta \psi_c = \frac14 \omega_{ab} \gamma^{ab} \psi_c +  {\omega_c}^a \psi_a$.] Because these constraints do not involve derivatives of the fields,  they generate proper gauge transformations with vanishing charge, even when the rotation parameter does not go to zero at infinity.

\subsection{Boundary conditions}
\subsubsection{Graviton}
We consider asymptotically flat spacetimes, which we describe in asymptotic Minkowskian coordinates.  The slices of constant time are asymptotic parallel hyperplanes. The spatial metric reads
\be
g_{ij} = \delta_{ij} + h_{ij}
\ee
where $h_{ij}$ decays as $r^{-1}$ in a manner that we shall specify below.  The extrinsic curvature ($\sim \pi^{ij}$) decays as $r^{-2}$.

We asymptotically freeze the freedom of performing arbitrary spatial rotations of the triads by imposing
\be
e^a_i = \delta^a_i + \frac 12 \delta^{aj} h_{ij} + O\left(\frac{1}{r^2}\right) \label{eq:GaugeF1}
\ee
In that gauge, we find that 
\be
\pi^{i}_a = 2 \delta_{aj} \pi^{ij} + O\left(\frac{1}{r^3}\right) \label{eq:GaugeF2}
\ee
The lowest order terms need not be rotation-gauge-fixed. 

Because the local Lorentz gauge freedom is fixed at infinity, Poincar\'e transformations need to be supplemented by local Lorentz transformations that bring one back to the time gauge $e^0_i =0$ and to (\ref{eq:GaugeF1}). More specifically, asymptotic boosts and rotations (characterized by vector fields $\xi^\rho$ such that $\d_\mu \xi^\rho \not=0$) acting through the standard Lie derivative on the local frames $\{e^\mu_\Delta \}$, induce terms that violate the gauge conditions  since $\mathcal{L}_{\xi^\rho} e^0_i \not=0$ and $\mathcal{L}_{\xi^\rho} e^a_i \not=0 $ (even to leading order).  These must be compensated by local Lorentz transformations of the local frames that bring one back to the chosen gauge.  Of course, this is automatically taken into account if one uses the Dirac bracket associated with the (partial) gauge fixing $e^0_k = 0$, $e^a_k = \delta^a_k+ $ subleading terms.

Since the local frames have been tied to the metric asymptotically, it is  only necessary to give the boundary conditions on the metric variables $h_{ij}$ and $\pi^{ij}$ in order to specify the boundary conditions on the spin-$2$ variables $(e^a_i, \pi^i_a)$.  But this is precisely what was achieved in \cite{Henneaux:2018hdj,Henneaux:2019yax}, from which we reproduce the boundary conditions. We start with the spatial metric.  One has
\be
 h_{ij} \equiv g_{ij} - \delta_{ij}  =   \frac{\xbar h_{ij}(\mathbf{n}^k)}{r} + O\left(\frac{1}{r^2}\right),  \label{eq:asymptgrav0a}
 \ee
where $\mathbf{n}^k$ is the unit normal to the sphere ($\mathbf{n}^i = \frac{x^i}{r}$), so that $\xbar h_{ij}$ is a function of the angles $\theta, \varphi$.  We decompose the coefficient $\xbar h_{ij}$ of the leading order into even and odd parts, 
\be
\xbar h_{ij}(\mathbf{n}^k) = (\xbar h_{ij})^{even}(\mathbf{n}^k) + (\xbar h_{ij})^{odd}(\mathbf{n}^k)
\ee
with
\begin{eqnarray}
&&(\xbar h_{ij})^{even} (-\mathbf{n}^k) = (\xbar h_{ij})^{even} (\mathbf{n}^k), \\
&& (\xbar h_{ij})^{odd} (-\mathbf{n}^k) = -(\xbar h_{ij})^{odd} (\mathbf{n}^k).
\end{eqnarray}
The even part is arbitrary.  Contrary to the strict parity conditions of \cite{Regge:1974zd}, where $(\xbar h_{ij})^{odd} (\mathbf{n}^k)$ was set to zero, we do allow a non-vanishing $(\xbar h_{ij})^{odd} (\mathbf{n}^k)$,  however.  But we impose that $(\xbar h_{ij})^{odd} (\mathbf{n}^k)$ should be generated by an improper gauge transformation parametrized by a vector that depends on the angles only (in order to preserve the $1/r $ decay of $h_{ij}$),
\begin{eqnarray}
&& (\xbar h_{ij})^{odd} (\mathbf{n}^k) = r U_{ij} \\
&& U_{ij} =
		\d_i \zeta_j + \d_j \zeta_i  = O(\frac{1}{r}),  \\
&&  \zeta^i = \zeta^i (\mathbf{n}^k) = O(1), \quad   \zeta^{i} (-\mathbf{n}^k) = \zeta^{i} (\mathbf{n}^k), \label{eq:asymptgrav0b}
\end{eqnarray}
for some vector $\zeta^i (\mathbf{n}^k)$  that may be assumed to be even since the odd parity component can be absorbed in a redefinition of $(\xbar h_{ij})^{even}$. 
Thus, the leading order ($O(1/r)$) of $h_{ij}$ is even up to an improper gauge transformation parametrized by $\zeta_i$. A twist is allowed in the parity conditions, given by an improper gauge transformation.

Similarly,   we allow the leading order ($O(1/r^2)$) of $\pi^{ij}$,
\be
\pi^{ij} = \frac{\xbar \pi^{ij}(\mathbf{n}^k)}{r^2} + O\left(\frac{1}{r^3}\right) \label{eq:asymptgrav0dd}
\ee
to have both an odd component $(\xbar \pi^{ij})^{odd}(\mathbf{n}^k)$ (which is the only component allowed in \cite{Regge:1974zd}) {\em and} an even component $(\xbar \pi^{ij})^{even}(\mathbf{n}^k)$ in $\xbar \pi^{ij}$,
\be
\xbar \pi^{ij}(\mathbf{n}^k) = (\xbar \pi^{ij})^{odd} (\mathbf{n}^k) + (\xbar \pi^{ij})^{even} (\mathbf{n}^k)
\ee
with
\begin{eqnarray}
&&  (\xbar \pi^{ij})^{odd} (-\mathbf{n}^k) = - (\xbar \pi^{ij})^{odd} (\mathbf{n}^k) \label{eq:asymptgrav0f} \\
&&  (\xbar \pi^{ij})^{even} (-\mathbf{n}^k) =  (\xbar \pi^{ij})^{even} (\mathbf{n}^k) \label{eq:asymptgrav0fF}
\end{eqnarray}
The odd component is unrestricted, but the even component must come from the transformation of $\pi^{ij}$ under diffeomorphisms that go to constants at infinity. At leading order, $ \pi^{ij}$ sees only the normal diffeomorphisms, which we denote by $V$, with $V$ of order one.   The transformation of $\pi^{ij}$ takes the form $\d^i \d^j V - \delta^{ij} \mathring \triangle V$ at leading order, where $\mathring \triangle \equiv \mathring \nabla ^i \mathring \nabla_i $ is the flat metric Laplacian.  The condition on $(\xbar \pi^{ij})^{even} (\mathbf{n}^k)$ is thus
\begin{eqnarray}
&& (\xbar \pi^{ij})^{even} (\mathbf{n}^k) =  r^2 V^{ij}  \label{eq:asymptgrav0d}\\
&& V^{ij} = \d^i \d^j V - \delta^{ij} \mathring \triangle V = O\left(\frac{1}{r^2}\right) \\
&&  V= V(\mathbf{n}^k) = O(1), \quad   V (-\mathbf{n}^k) = V (\mathbf{n}^k) \label{eq:asymptgrav0e} 
\end{eqnarray}
for some $V$ that may be assumed to be even since the odd parity component can be absorbed in a redefinition of $(\xbar \pi^{ij})^{odd}$.  

Because the transformations linearize at infinity, the finite forms of the improper gauge transformations $U_{ij}$ and $V^{ij}$ coincide with their infinitesimal forms.  We can therefore assume that $\zeta^i$ and $V$ are finite, and not just infinitesimal, in the above formulas.

These parity conditions on the leading orders of the metric and the extrinsic curvature imply that the leading terms in the expansion of the electric and magnetic components of the Weyl tensor, which are invariant under proper and improper gauge transformations, be strictly even in cartesian coordinates.  Together with appropriate parity conditions on a BMS invariant metric function related to the mass  and its BMS invariant conjugate related to the linear momentum, spelled out in detail in \cite{Henneaux:2018hdj,Henneaux:2019yax},  these strict parity conditions on the Weyl tensor imply the above parity conditions with a twist on the spatial metric and its conjugate momentum.

The parity conditions with an improper gauge twist are the analogs of the generalized parity conditions imposed on $1$-forms and $2$-forms
\cite{Henneaux:2018gfi,Henneaux:2018mgn,Henneaux:2019yqq}.

As explained in \cite{Henneaux:2018hdj,Henneaux:2019yax}, it turns out that the metric variables must be subject to one extra condition, which is
that the mixed radial-angular components $h_{rA}$  of the metric perturbation, which is potentially of order $O(1)$, should actually decrease one power of $r^{-1}$ faster, i.e., 
\be
h_{rA} = O\left(\frac{1}{r}\right).  \label{eq:AsymCondhrA}
\ee
This implies in particular 
\begin{eqnarray} 
&& \zeta_i = \d_i \tilde U, \quad \tilde U = r \xbar U,   \\ 
&& \xbar U = \xbar U(\mathbf{n}^k) = O(1), \\
&& \xbar U (-\mathbf{n}^k) = -\xbar U (\mathbf{n}^k) \label{eq:asymptgrav2aA}
\end{eqnarray}
for some function $\xbar U$ of the angles.  The condition (\ref{eq:AsymCondhrA}) plays a crucial role for the emergence of the BMS group, and we refer to \cite{Henneaux:2018hdj,Henneaux:2019yax} for more information.

\subsubsection{Gravitino}
We now turn to the gravitino.  At least three requirements must be fulfilled by the searched-for boundary conditions. (i) They should make the action finite. (ii) They should be invariant under global supersymmetry, characterized by a supersymmetry parameter $\varepsilon$ that goes to a constant spinor at infinity.  (iii) They should make the surface integrals appearing in the supersymmetry generators finite.

Since the fermionic kinetic term is quadratic in $\psi_k$, one way to make it finite is to take
\be
\psi_k = \frac{\chi_k(\mathbf{n}^k)}{r^2} + O\left(\frac{1}{r^3} \right) \label{eq:BCFermion}
\ee
where $\chi$ is a function of the angles to which we impose tentatively no parity condition since the integrand of the fermionic kinetic term is then of order $\sim 1/r^4$ and yields a convergent integral without the need for cancellations dictated by parity conditions. 

We now check that actually, (\ref{eq:BCFermion}) successfully meets all the requirements.

To leading order, the supersymmetry transformations acting on the canonical variables are easily verified to be (see e.g \cite{Bunster:2012jp}, noting that $\varepsilon^{\textrm{here}}  = 4 \epsilon^{\textrm{there}} $), 
\begin{eqnarray}
&& \delta_\varepsilon h_{ij} = \frac12 i \xbar \varepsilon \gamma_{(i} \psi_{j)} \\
&& \delta_\varepsilon \pi^{ij} = - \frac{i}{8} \xbar \varepsilon \gamma_0 \d^i \psi^j - \frac{i}{8}  \epsilon^{irs} \xbar \varepsilon  \gamma^j \gamma_5 \d_r \psi_s \nonumber \\
&& \hspace{1cm} + \frac{i}{8} \delta^{ij} \xbar \varepsilon \gamma_0 \partial^k \psi_k + (i \leftrightarrow j), \\
&& \delta_\varepsilon \psi_i =  \d_i \varepsilon - \frac14 \partial_r h_{is} \gamma^{rs} \varepsilon + \frac12 K_{is} \gamma^{0} \gamma^{s} \varepsilon \label{eq:TRGravitino}
\end{eqnarray}
with $ \varepsilon = \varepsilon_0 + O(1/r)$ and $K_{ij} = - \pi_{ij} + \frac12 \pi \delta_{ij}$ (asymptotically).  Here, $\varepsilon_0$ is a constant spinor. The right-hand side of the first line is of order $O(1/r^2)$ and so preserves the boundary conditions on $h_{ij}$ since it does not affect the leading, more tricky $O(1/r)$-part which is subject to non trivial parity conditions.  Furthermore, one has (in polar coordinates) $\delta_\varepsilon h_{rA} = O(r^{-1})$ so that (\ref{eq:AsymCondhrA}) is clearly preserved. Similarly, the leading order of $\pi^{ij}$ is unaffected and it is only the subleading $O(1/r^3)$-term, subject to no parity condition, that transforms under supersymmetry.   Thus, the boundary conditions on the bosonic fields are preserved under global supersymmetry.
The same also holds for the gravitino field, since the right-hand side of (\ref{eq:TRGravitino}) is evidently of order $O(1/r^2)$.

Finally, the surface integral appearing in the supersymmetry generator 
\be
S_\varepsilon = i \int d^3 x \varepsilon^T {\mathcal S} + B_{\textrm{Susy}}
\ee
is determined by the requirement that $S_\varepsilon$
with asymptotically constant $\varepsilon$ should be well-defined, which yields (in cartesian coordinates)\cite{Teitelboim:1977hc,Deser:1977hu}
\be
\delta B_{\textrm{Susy}} = - i \oint_{S^\infty} d^2 S_m  \varepsilon^T \gamma^{mn} \delta \psi_n.  \label{eq:VarSusyCharge}
\ee
Here,  the surface element $d^2 S_l$ is equal to $r^2 n_l d^2S$ where $d^2 S$ is the surface element on the unit sphere, and so $d^2 S_l v^l = r^2 v^r \sin \theta d \theta d \varphi$ in standard polar coordinates, where $v^r$ is the radial component of the vector $v^l$, $v^r \equiv \mathbf{n \cdot v}$.   It is clear from the asymptotic behaviour of the various quantities entering (\ref{eq:VarSusyCharge}) that $\delta B_{\textrm{Susy}}$ is finite.  Furthermore, it is clearly integrable because one can replace $\gamma^{mn}$ by the field-independent flat space expression and pull the $\delta$ out of the integral. One has, since only the constant piece $\varepsilon_0$ of the supersymmetry parameter contributes to the surface integral, 
\be
B_{\textrm{Susy}} = - i  \varepsilon^T_0 \oint_{S^\infty} d^2 S_m   \gamma^{mn} \psi_n,
\ee
an expression that can be transformed to
$$
B_{\textrm{Susy}} = - i  \varepsilon^T_0 \oint_{S^\infty} d^2 S_m   \gamma^{mn}_{FS} \frac{\chi_n}{r^2}
$$
to emphasize that the leading flat space piece $\gamma^{mn}_{FS}$ of $\gamma^{mn}$ and the leading piece $\chi_n$ of the gravitino field are the only relevant ones in the surface integral. 

The supersymmetry generator is consequently well defined and we have thus verified that (\ref{eq:BCFermion}) fulfills all three requirements listed above. 

There is one more technical condition that must be imposed on the asymptotic behaviour \cite{Henneaux:2018hdj,Henneaux:2019yax}.   For generic decays, the
constraints $\mathcal H $ and $\mathcal H_i $ typically
behave as $r^{-3}$ in cartesian coordinates.  We require them to go to zero two powers of $r^{-1}$ faster, i.e., 
\begin{eqnarray}
&& \mathcal H= O(r^{-5}), \quad \mathcal H_i = O(r^{-5})  \nonumber \\
&& \hspace{.5cm}  \hbox{(in cartesian coordinates)}. \label{eq:asymptgrav1}
\end{eqnarray}
With the fall-off of the gravitino field, the terms $F_1$ and $F_2$ in $\mathcal H $ and $\mathcal H_i $ automatically decay as $O(r^{-5})$, and so this condition, which we impose, is a restriction on the graviton field only.   The algebra of the constraints, given in \cite{Teitelboim:1977fs}, guarantees that this fall-off is preserved under the transformations generated by the constraints.

Our complete set of boundary conditions is thus (\ref{eq:asymptgrav0a})--(\ref{eq:asymptgrav0b}),  (\ref{eq:asymptgrav0dd})--(\ref{eq:asymptgrav0e}), (\ref{eq:AsymCondhrA}),  (\ref{eq:BCFermion}) and (\ref{eq:asymptgrav1}), with the tetrad variables asymptotically related to the metric  variables as in (\ref{eq:GaugeF1}) and (\ref{eq:GaugeF2}).

\section{Asymptotic symmetries}
\label{sec:AsSymm}
\setcounter{equation}{0}

The boundary conditions are invariant under a larger set of transformations than the super-Poincar\'e algebra.

\subsection{Lorentz transformations}

The homogeneous Lorentz transformations are described by surface deformations $\xi$, $\xi^i$ that behave asymptotically as 
\begin{eqnarray}
&& \xi = b_i x^i + C^{(b)}(\mathbf{n}) + O\left(r^{-1}\right) \label{eq:AsLorentz1}\\
&&  \xi^i ={ b^i}_j x^j + C^{(b)i} (\mathbf{n})+ O\left(r^{-1}\right) \label{eq:AsLorentz2}
\end{eqnarray}
where $b_i$ and $b_{ij} = -b_{ji}$ are arbitrary constants.    The constants $b_i$ parametrize the Lorentz
boosts (the corresponding term $- b^i x^0$ in $\xi^i$ can be absorbed in a spatial translation
at any given time and will be discussed with the translations), whereas the antisymmetric constants $b_{ij} = -b_{ji}$
parametrize the spatial rotations.  The homogeneous Lorentz transformations blow up linearly in $r$. 
The subleading $O(1)$ terms $C^{(b)}(\mathbf{n})$ and $C^{(b)}_i (\mathbf{n})$ are ``correcting terms''
that appear only when the transformation involves a boost ($b_i \not=0$) \cite{Henneaux:2018hdj,Henneaux:2019yax}. As explained there, they are necessary for integrability of the boost charges and for maintaining the condition $\xbar h_{Ar} = 0$.  These terms are linear in the boost parameters so that they vanish when $b_i= 0$.  The inclusion of the fermions does not modify the discussion of these terms and so we refer to \cite{Henneaux:2018hdj,Henneaux:2019yax} for more information.

The invariance of the boundary conditions on the bosonic fields have been verified in \cite{Henneaux:2018hdj,Henneaux:2019yax}. The extra contributions proportional to the fermions, which appear in the boost variations of the bosonic fields, do not invalidate this result since these are subleading and of order $O(r^{-3})$.  The boundary conditions on the fermionic field are also invariant, because one has $\delta_{\textrm{boost}} \psi \sim \xi \d \psi$ (to leading order), and this is $O(r^{-2})$ as requested. 

\subsection{Translations and BMS supertranslations}
The boundary conditions are also invariant under translations and BMS supertranslations,
\begin{eqnarray}
&& \xi =  T(\mathbf{n})  + O\left(r^{-1}\right) \label{eq:AsInhom1}\\
&&  \xi^i = W^i(\mathbf{n}) + O\left(r^{-1}\right) \label{eq:AsInhom2} \\
&& W_i(\mathbf{n}) =  \d_i (r W(\mathbf{n})) \label{eq:AsInhom3}
\end{eqnarray}
where
$T(\mathbf{n})$ and $W(\mathbf{n})$ are arbitrary functions on the unit
sphere.     The zero modes $a_0$ and $a_0^i$ of $T$
and $W^i$ are standard translations.  In a spherical harmonics expansion of $T(x^A)$ and $W(x^B)$ ($X^A \equiv$ coordinates on the unit sphere), this corresponds to the choices $T(x^B) \sim a_0 Y^{0}_{0} $ and $W(x^B)  \sim a_0^m Y^1_m$ but higher spherical harmonics are allowed.  These higher harmonics yield ``BMS supertranslations'' and lead to an infinite-dimensional extension of the homogeneous Lorentz group.  As in the pure bosonic case, the odd part of $T$ and the even part of $W$ define proper gauge transformations that do not change the physical state of the system (see below).  Only the even part of $T$ and the odd part of $W$ define improper gauge transformations with non trivial action on the system.  These combine furthermore to yield a function on the $2$-sphere with both even and odd parts, which corresponds to the standard parametrization of the BMS supertranslations \cite{Henneaux:2018hdj,Henneaux:2019yax,Troessaert:2017jcm}. 

The invariance of the boundary conditions for the graviton field was checked in \cite{Henneaux:2018hdj,Henneaux:2019yax} and is unaffected by the fermion contributions, which are subleading. The invariance of the boundary conditions for the gravitino field is also immediate.

\subsection{Supersymmetry}

We now turn to the fermionic symmetries.  It is clear that the boundary conditions are invariant under supersymmetry transformations that behave asymptotically as
\be
\varepsilon = \varepsilon_0 + O(1/r)
\ee
where $\varepsilon_0$ is a {\it constant} spinor.   No angular dependence is allowed in $\varepsilon_0$, since this would lead to unwanted $O(r^{-1})$ terms in 
$\delta_\varepsilon \psi_k$ through 
$$\d_k \varepsilon_0 = \frac{\partial \varepsilon_0} {\partial \theta} \frac{\partial \theta} {\partial x^k} +  \frac{\partial \varepsilon_0} {\partial \varphi} \frac{\partial \varphi} {\partial x^k}, \quad \frac{\partial \theta} {\partial x^k}, \frac{\partial \varphi} {\partial x^k} \sim \frac{1}{r} ,
$$ 
which could not be compensated.  Hence, the global part of the supersymmetry transformations involve only four independent real fermionic parameters and is not infinite-dimensional.

\subsection{Generators}

The generators of the above transformations are combinations of the constraints plus a surface term,
\begin{eqnarray}
	P_{\xi,\xi^i, \varepsilon}[g_{ij}, \pi^{ij}, \psi_k] &=& \int d^3x \, \left(\xi \mathcal H + \xi^i \mathcal H_i
	+ i \varepsilon^T \mathcal{S} \right) \nonumber \\
	&& + \mathcal B_{\xi,\xi^i, \varepsilon} \label{eq:bms4sugra}
\end{eqnarray}
where the boundary term $\mathcal B_{\xi,\xi^i, \varepsilon}$ is determined by the method of \cite{Regge:1974zd}, i.e., must be such that the exterior derivative $d_V P_{\xi,\xi^i, \varepsilon}$ of $P_{\xi,\xi^i, \varepsilon}$ in field space (with $(\xi, \xi^k, \varepsilon)$ given above) reduces to a bulk integral involving only undifferentiated field variations $d_V e^a_i$, $d_V \pi_a^i$, $d_V \psi_k$.  Differently put, the variation $d_V \mathcal B_{\xi,\xi^i, \varepsilon}$ of the surface term  must cancel the boundary terms generated from $d_V \int d^3x \, \left(\xi \mathcal H + \xi^i \mathcal H_i + i \varepsilon^T \mathcal{S}
	\right)$ through the integrations by parts necessary to bring $d_V P_{\xi,\xi^i, \varepsilon}$ to the appropriate bulk form. This is equivalent to requesting that the transformations of the fields under super-BMS transformations be canonical transformations, i.e., leave the symplectic form invariant, $d_V i_\xi \Omega = 0$ ($\Leftrightarrow i_\xi \Omega = - d_V P_{\xi,\xi^i, \varepsilon}$, see \cite{Henneaux:2018gfi}).  Note in particular that the fermionic kinetic term is invariant under boosts without need to add a surface term, so that the symplectic form is pure bulk.

The computation of the surface term accompanying the bosonic transformations (homogeneous Lorentz transformations, translations and BMS supertranslations) has been carried out in \cite{Henneaux:2018hdj,Henneaux:2019yax} for pure gravity.   Now, the terms in the derivatives of the fermionic field in $\mathcal H $ and $ \mathcal H_i$, which could potentially contribute to the surface integrals, have the form $\psi \d \psi$.  The variation of these terms leads to surface integrals of the form $\oint d^2 S \xi \psi d_V \psi$, which goes to zero like $r^2 r r^{-2} r^{-2} \sim r^{-1}$.  Hence the fermions do not contribute to the surface integrals of the bosonic charges, which can be taken unchanged from \cite{Henneaux:2018hdj,Henneaux:2019yax}.  Adding the surface integral for supersymmetry transformations derived above, one therefore gets
\be
 \mathcal B_{\xi,\xi^i, \varepsilon} = b_i K^i 
     + \frac 12 b_{mn} M^{mn} + \mathcal B_{\{T, W\}} + i \varepsilon^T_0 \mathcal B_{\textrm{Susy}}
 \ee
 with
 \begin{eqnarray}
b_i K^i  &=& \oint d^2x \, \Big \{ b\,
	\sqrt{\xbar \gamma} \Big( 2  k^{(2)} 
	 + \xbar k^2 +
	\xbar k^A_B \xbar k^B_A - 6 \xbar\lambda \, \xbar k\Big)  \nonumber \\ &&  +b\frac{2}{\sqrt{\xbar\gamma}} \xbar \gamma_{AB}
    \xbar\pi^{rA}\xbar\pi^{rB}\Big\} , \quad b(\mathbf{n}) = b_i \frac{x^i}{r}  \label{eq:BoostCharge}
 \end{eqnarray}
(boosts), 
\be
 M^{mn} = 4\oint x^{[n}\Big(\ \Pi^{(3) m]l} + \delta^{m]p}h^{(1)}_{pk}  \Pi^{(2) kl} \Big) d^2 S_l
 \ee
 (spatial rotations), 
 \begin{eqnarray}
 \mathcal B_{\{T, W\}} &=& \oint \mathring G^{ijkl} \, T^{even}  \,  \Big((h^{ (1)}_{ij})^{even}\Big)_{, k}  d^2 S_l  \hspace{1cm}  \label{eq:BMSSup}\\
 && + \oint \left(2 \d_k( W^{odd})   (\Pi^{ (2) kl} )^{odd} \right) d^2 S_l \hspace{1cm} 
 \end{eqnarray}
 (translations and BMS supertranslations),
 \be
 \mathcal B_{\textrm{Susy}} = -  i \oint d^2 S_m \gamma^{mn} \psi_n
 \ee
 (supersymmetry transformations).
 
 A few words of explanation are needed to understand these formulas.  
 \begin{itemize}
 \item {\bf Boosts:}  the boost surface integrals have been written in polar coordinates as this turns out to be more convenient.  Each term in the integral (\ref{eq:BoostCharge}) depends only on the angles $x^A$ ($A= 1,2$) and does not involve the radial coordinate $r$.  One has $d^2 x = dx^1 dx^2 = d \theta d \varphi$ (if one uses  polar coordinates) and $\xbar \gamma _{AB} $ is the round metric on the unit sphere, $\xbar \gamma_{AB} dx^A dx^B = d \theta^2 + \sin^2 \theta d \varphi^2$, with determinant $\xbar \gamma = \sin^2 \theta$.  The function $\xbar \lambda$ is defined through the expansion of $g_{rr}$,
  \be
 g_{rr} =  1 +  \frac{2 \xbar \lambda}{r} + O(r^{-2})
 \ee
Similarly, the functions $k$ are introduced through an expansion of the extrinsic curvature $K_{AB}$ of the spheres of constant $r$ (in the constant time hypersurfaces),
\begin{equation}
	 K^A_B = - \frac{\delta^A_B}{r}  + \frac{\xbar k^A_B}{r^2} +
	 \frac{{k^{(2)}}^A_B}{r^3} + O\left(\frac{1}{r^{4}}\right)
\end{equation}
and $\xbar k = \delta_A^B \, \xbar k^A_B$.  We have also expanded  the momenta $\pi^{ij}$ (which are densities) in polar coordinates,
\begin{eqnarray}
		\pi^{rr} &=& \xbar \pi^{rr} + \frac{1}{r} \pi^{(2)rr} +
		O(r^{-2}),\\
		\pi^{rA} & = & \frac{1}{r} \xbar \pi^{rA} + \frac{1}{r^2} \pi^{(2)rA} +
	O(r^{-3}),\\
	\pi^{AB} & = & \frac{1}{r^2} \xbar \pi^{AB} + \frac{1}{r^3} \pi^{(2)AB}+
	O(r^{-4}).
\end{eqnarray}
Note the presence of nonlinear terms in the deviation from the flat metric.
 \item {\bf Spatial rotations:} The other charges have been written in cartesian coordinates. The individual terms in the integrand involve $r$ and one can check that the integrals are finite.  The terms $\Pi^{(k) ml}$ is the term of order $r^{-k}$ in the expansion of $\pi^{ml}$, e.g.,
 $$
 \Pi^{(2) ml} = \frac{\xbar \pi^{ml}(\mathbf{n})}{r^2}
 $$ 
 and similarly, The terms $h^{(k)}_{ ml}$ is the term of order $r^{-k}$ in the expansion of $h_{ml}$, e.g.,
 $$
 h^{(1)}_{ ml} = \frac{\xbar h_{ml}(\mathbf{n})}{r}.
 $$
  The integrals yielding the angular momentum $M^{mn}$ are finite since the powers of $r$ cancel, explicitly one gets $r$ (for $x^n$) $\times  r^{-3}$ (for $\Pi^{(3)}$) $\times r^2$ (for $d^2 S_l$) and $r r^{-1} r^{-2} r^2$ for the other term. 
 
 One can rewrite the angular momentum in spherical coordinates \cite{Henneaux:2018hdj,Henneaux:2019yax},
 \begin{eqnarray}
	&& \frac 12 b_{mn} M^{mn} = \nonumber \\ &&\qquad \qquad \oint d^2x \Big
	\{  
	Y^A \Big(4\xbar k_{AB} \xbar \pi^{rB} - 4 \xbar \lambda\xbar \gamma_{AB}
	\xbar\pi^{rB} \nonumber \\ && \qquad \qquad \qquad  \quad + 2 \xbar \gamma_{AB}
	\pi^{(2)rB}\Big) \Big\}
\end{eqnarray}
where $Y^A = \frac12 b^{mn} Y^A_{mn}$ are the rotation Killing vectors, which are tangent on the sphere and have only angular components.
 
  \item {\bf Translations and BMS supertranslations:} In (\ref{eq:BMSSup}), $\mathring G^{ijkl}$ is the De Witt  supermetric for the flat metric in cartesian coordinates $\delta_{ij}$,
\be
\mathring G^{ijkl}  =  \frac12 (\delta^{ik} \delta^{jl} + \delta^{il} \delta^{jk}) - \delta^{ij} \delta^{kl} .
\ee 
One important feature of the formula giving the BMS supertranslation charges is that it involves only $T^{even}$ and $\tilde W^{odd}$, as announced above.  The opposite parity components $T^{odd}$ and $\tilde W^{even}$ drop from the formulas and define proper gauge transformations that can be factored out. Even after this quotient is taken, there exists an infinite number of improper gauge symmetries with non trivial action.  

There is an infinite number of conserved charges.  These do not receive fermionic contributions as  the gravitino field decays too fast at infinity.  In particular, the energy, corresponding to $T = 1$, $\tilde W = 0$ is unchanged and coincides with the ADM energy \cite{Arnowitt:1962hi,Teitelboim:1977hc}.

In polar coordinates,
these charges read \cite{Henneaux:2018hdj,Henneaux:2019yax},
\begin{eqnarray}
 \mathcal B^{grav}_{\{T, W\}} &= &
 \oint d^2x \Big
	\{   2 W^{odd} \Big( \xbar \pi^{rr} -  \xbar \pi^A_A\Big)  \nonumber \\
	&&\qquad \qquad + 4 \, T^{even} \sqrt{\xbar\gamma}\,  \xbar
	\lambda \Big\}
\end{eqnarray}
  \item {\bf Supersymmetry:} Only the constant part $\varepsilon_0$ appears in the surface integral at infinity.  The subleading terms ($O(r^{-1})$) are proper gauge transformations that do not change the physical state of the system.  There are only four non trivial global supersymmetry transformations.
 \end{itemize}

\subsection{Lagrange multipliers}
The Lagrange multipliers $N$,  $N^k$ and $\psi_0$  must be chosen so that the dynamical evolution preserves the boundary conditions.  This means that they can be taken to parametrize a generic asymptotic symmetry.  It is customary to take:
\begin{equation}
	\label{eq:asymptgravII}
	N = 1 + O(r^{-1}), \quad N^k = O(r^{-1}), \quad  \psi_0 = O(r^{-1}).
\end{equation}
This corresponds to slicings by hypersurfaces that become asymptotically parallel hyperplanes, with no supersymmetry transformation performed as one marches on. Imposing these boundary conditions on the lapse and the shift implies that we have to add to the action  the ADM energy, i.e, 
\begin{equation}
B_{\infty} = \mathcal B_{\{1, 0, 0\}} = \oint d^2x
	\sqrt{\xbar \gamma} \,4 \xbar \lambda.
	\label{eq:BounTermAction0}
\end{equation} 	

The internal, local rotation Lagrange multipliers $\lambda_{ab}$ are asymptotically fixed by the gauge condition fixing the triads $e^a_i$ in terms of the metric and does not come with a surface term anyway.

\section{Asymptotic symmetry algebra}
\label{sec:AsymptoticAlgebra}
\setcounter{equation}{0}

The algebra of the generators is easily evaluated to be:
\begin{eqnarray}
	&& \Big\{P_{\xi_1, \xi^i_1, \varepsilon_1}[g_{ij}, \pi^{ij}, \psi_k], P_{\xi_2, \xi^i_2, \varepsilon_2}[g_{ij}, \pi^{ij}, \psi_k]\Big\} \nonumber \\ && \qquad \qquad =
	P_{\hat\xi, \hat\xi^i,\hat\varepsilon}[g_{ij}, \pi^{ij}, \psi_k],  \label{eq:PBAlgebraSUSY}
\end{eqnarray}
where the triplet ($\hat \xi, \hat\xi^i,\hat\varepsilon$) generates an asymptotic symmetry with the following asymptotic parameters
\begin{flalign}
	\label{eq:hamilbmsI}
	\hat Y^A & = Y^B_1\d_B Y_2^A + \xbar \gamma^{AB} b_1\d_B b_2 - (1
	\leftrightarrow 2),\\
	\label{eq:hamilbmsII}
	\hat b & = Y^B_1\d_B b_2 - (1 \leftrightarrow 2),\\
	\label{eq:hamilbmsIII}
	\hat T & = \frac{i}{8} \varepsilon_{0,1}^T \varepsilon_{0,2} +Y_1^A\d_A T_2 - 3 b_1 W_2  \nonumber \\ & - \d_A b_1 \xbar D^A W_2 - b_1
	\xbar D_A\xbar D^A W_2 - (1 \leftrightarrow 2),\\
	\label{eq:hamilbmsIV}
	\hat W & = \frac{i}{8} \varepsilon_{0,1}^T \gamma_0 \gamma_i n^i\varepsilon_{0,2} + Y_1^A \d_A W_2 - b_1T_2 \nonumber \\ & \qquad - (1 \leftrightarrow 2), \\
	\hat \varepsilon_0 & = \frac14 b^1_{mn} \gamma^{mn} \varepsilon_{0,2} + \frac12 b^1_i \gamma^{0i} \varepsilon_{0,2} - (1 \leftrightarrow 2) \label{eq:TransfSusyCharge}
\end{flalign}
(the factor $\frac18$ in front of the terms bilinear in the fermionic parameters is due to our normalization conventions).

To establish this bracket algebra, one computes the commutator of the corresponding known transformations and uses the general theorems that guarantee that the canonical generators realize the algebra of the transformations up to a central charge.  So, once one knows the algebra of the transformations,  one only needs to compute the central charge, which is easily seen to vanish  here if one adjusts the value of the canonical generators to zero on the Minkowski solution (with $\psi_k = 0$) -- as implicitly done above -- since the variation of the charges is then zero.  See \cite{Brown:1986ed,Brown:1986nw} for more information.  The procedure is illustrated in Appendix \ref{sec:appSusyBra}.

The Lorentz transformations mix $T^{even}$ with $W^{odd}$, and $T^{odd}$ with
$W^{even}$ since the boost parameters $b$ are odd functions on the sphere.  Setting the proper gauge multiplet $(T^{odd}, W^{even})$ equal to zero is consistent with Lorentz invariance.

The bracket of two supersymmetry transformations generically yields non trivial $T$'s and $W$'s.  We see, however,  that only the zero mode of the resulting $T$ and the first harmonics ($\sim n^i$) of the resulting $W$ are non zero.  These correspond to ordinary time and space translations. The anticommutator of two supersymmetry transformations does not yield a non trivial BMS supertranslation, in agreement with \cite{Awada:1985by}.

In fact, the above superalgebra is exactly the super-BMS algebra of \cite{Awada:1985by}.
This is because the bosonic part of the algebra is the standard $BMS_4$ algebra expressed in an unfamiliar parametrization, as shown in \cite{Henneaux:2018cst,Henneaux:2018hdj} (using the results of \cite{Troessaert:2017jcm}).

\section{Extended Supergravity}
\label{sec:ExtenstedSUSY}
\setcounter{equation}{0}

The above analysis can be straightforwardly generalized to extended supergravity models.  We consider the $N=2$ case \cite{Ferrara:1976fu,Ferrara:1976iq,Fradkin:1979cw,deWit:1979xpv,deWit:1979dzm}, which illustrates the main points.

In addition to the graviton and the two gravitini, the $N=2$ supergravity model contains a photon.  We adopt for the corresponding vector field $A_i$ and its conjugate momentum $\pi^i$ the parity conditions with an improper gauge twist of \cite{Henneaux:2018hdj,Henneaux:2018gfi}, i.e.,
\begin{eqnarray}
&& A_i = \frac{\xbar A_i (\mathbf{n}^k)}{r} + O\left(\frac{1}{r^2}\right),  \\
&& \xbar A_i (\mathbf{n}^k)=  \xbar A_i^{even}(\mathbf{n}^k)+ \xbar A_i^{odd}(\mathbf{n}^k), \\
&& \xbar A_i^{odd}(\mathbf{n}^k) = r \partial_i \xbar \Phi (\mathbf{n}^k) , \\
&& \pi^i = \frac{\xbar \pi^i(\mathbf{n}^k)}{r^2} + O\left(\frac{1}{r^3}\right)\\
&& \xbar \pi^i(\mathbf{n}^k) = {\xbar \pi^i}^{odd}(\mathbf{n}^k).
\end{eqnarray}
The even part of the leading order of the vector potential is arbitrary but its odd  part is given by an improper gauge transformation, i.e., the gradient of a function $\xbar \Phi$ depending on the angles only and which can assumed to be even,
 \be
 \xbar \Phi (-\mathbf{n}^k) = \xbar \Phi (\mathbf{n}^k).
 \ee
 The odd part of the leading order of the conjugate momentum is arbitrary but its even part (which is invariant under proper and improper gauge transformations) vanishes,
 \be
 {\xbar \pi^i}^{even}(\mathbf{n}^k) = 0.
 \ee
 These boundary conditions are equivalent to the requirement that the leading orders of the electric and magnetic fields be strictly odd (in cartesian coordinates).
 
In addition to these variables, there is a boundary degree of freedom at infinity, denoted by $\xbar \Psi (\mathbf{n}^k)$,  which enters the kinetic term of the action through a surface term,
\be
S^{\textrm{em}} \sim \int dt \Big[\int d^3 x \pi^i \d_t A_i - \int_{S^\infty} d^2x \xbar A_r \d_t \xbar \Psi \Big]+ \textrm{``more''}
\ee
where the extra $\textrm{``more''}$ terms can be found in \cite{Henneaux:2018hdj,Henneaux:2018gfi}.

The electromagnetic action is invariant under the following gauge transformations
\begin{equation}
	\delta_{\phi,\mu} A_i = \d_i \phi,  \quad 
	\delta_{\phi,\mu} \pi^i = 0, \quad 
	\delta_{\phi,\mu} \xbar \Psi = \xbar \mu, 
\end{equation}
with
\begin{equation}
	\phi = \xbar \phi(\mathbf{n}^k)  + \frac 1 r \phi^{(1)} + O(r^{-2}),
	\quad \xbar \mu = \xbar \mu(\mathbf{n}^k) 
\end{equation}
The canonical generator of these transformations is given by
\begin{equation}
	\label{eq:generatorU1largeI}
	G_{\phi, \mu} = \textrm{``Bulk''} +
	\oint d^2x (\xbar \phi\, \xbar \pi^r - \sqrt{\xbar\gamma}\, \xbar \mu
	\xbar A_r)
\end{equation}
where the  $\textrm{``Bulk''} $ term is a linear combination of the Gauss constraint and of the constraint that expresses that the bulk extension of $\xbar \Psi$ is pure gauge. 
Only the transformations for which $\xbar\phi$ is even or $\xbar\mu$ is odd are improper gauge transformations since the radial component $\xbar \pi^r$ is even, while  $\xbar A_r$ is odd.  These even and odd functions combine to yield the angle-dependent $u(1)$ symmetry of electromagnetism, in perfect agreement with the null infinity analysis.  We refer to  \cite{Henneaux:2018hdj,Henneaux:2018gfi} for the details.

We denote the two gravitini by $\psi_k^{\mathfrak{B}}$ and the two supersymmetry parameters by $\varepsilon^{\mathfrak{B}}$ ($\mathfrak{A}, \mathfrak{B}, \mathfrak{C} = 1,2$). The metric and the Levi-Civita tensor in internal space are respectively $\delta_{\mathfrak{A} \mathfrak{B}}$ and $\epsilon_{\mathfrak{A} \mathfrak{B}}$.   We assume
\be
\psi_k^{\mathfrak{B}} = \frac{\chi_k^{\mathfrak{B}}(\mathbf{n}^k)}{r^2} + O\left(\frac{1}{r^3}\right) \label{eq:DecayGravitini}
\ee
as before.
To discuss the generalization of the $N=1$ theory to the $N=2$ theory, we need to show that the boundary conditions on the vector variables are preserved, and that the new terms appearing in the supersymmetry variations of the gravitini are also asymptotically acceptable.  Now, one has (to leading order)
\be
\delta_{\varepsilon} A_i \sim i \delta_{\mathfrak{B}\mathfrak{C}}\xbar \varepsilon^{\mathfrak{B}}  \psi_i^{\mathfrak{C}}, \quad
\delta_{\varepsilon} \pi^i \sim i\delta_{\mathfrak{B}\mathfrak{C}}\epsilon^{ijk}\xbar \varepsilon^{\mathfrak{B}} \gamma_5  \d_j\psi_k^{\mathfrak{C}}
\ee
and $\delta_{\varepsilon}  \xbar \Psi =0$, so that again, given the fast $1/r^2$ decay of the gravitini, these transformations only affect the subleading terms in $A_i$ and $\pi^i$ and manifestly preserve the boundary conditions.  Similarly, the transformation rule of the gravitini involves new terms of the form 
\be
\delta^{\textrm{extra}}_{\varepsilon} \psi_k^{\mathfrak{A}} \sim  \frac14 (\pi^m - \beta^m \gamma_5)(\delta_{km} \gamma_0 + \epsilon_{krm} \gamma_5 \gamma^r)  \varepsilon^{\mathfrak{A}}  
\ee
 (plus subleading terms), which are of order $O(r^{-2})$ and preserve therefore (\ref{eq:DecayGravitini}).   Here, $\beta^m$ is the magnetic field, $\beta^m = \frac12 \epsilon^{mij}F_{ij}$.  Thus, all boundary conditions are preserved.

Consequently, the action is invariant under the following improper gauge symmetries:
\begin{itemize}
\item BMS transformations (infinite-dimensional);
\item Angle-dependent $u(1)$ gauge symmetries (infinite-dimensional);
\item Extended supersymmetry, parametrized by two constant real spinors $\varepsilon_0^{\mathfrak{B}}$ (8-dimensional);
\item $su(2)$ transformations of the spinors (3-dimensional) \cite{Ferrara:1976iq}.
\end{itemize}
The extended $N=2$ super-BMS algebra contains again only a finite number of fermionic generators. 

The Poisson bracket algebra of the infinite-dimensional BMS and angle-dependent $u(1)$ transformations has been given in \cite{Henneaux:2018hdj,Henneaux:2018gfi}.  Accordingly, we only write here the bracket of the supersymmetry generators 
\begin{eqnarray}
S_{\varepsilon^{\mathfrak{B}}} &=& i \int d^3x (\varepsilon^{\mathfrak{B}})^T \mathcal{S}_{\mathfrak{B}} \nonumber \\
&& -  i (\varepsilon^{\mathfrak{B}}_0)^T \delta_{\mathfrak{B}\mathfrak{C}}\oint d^2 S_m \gamma^{mn} \psi^{\mathfrak{C}}_n
\end{eqnarray}
(where $\mathcal{S}_{\mathfrak{B}} \approx 0$ are the fermionic constraints) among themselves. These generators are Lorentz spinors transforming in the $\mathbf{2}$ of $su(2)$. Their brackets are 
\be
	 \Big\{S_{\varepsilon_1^{\mathfrak{A}}}, S_{\varepsilon_2^{\mathfrak{A}}}\Big\} 
	=
	P_{\hat\xi, \hat\xi^i}+G_{\hat{\phi}, \hat{\mu}} +  C_{\varepsilon_1^{\mathfrak{A}}, \varepsilon_1^{\mathfrak{A}}},
\ee
where the parameters ($\hat \xi, \hat\xi^i,\hat\phi, \hat \mu$) of the asymptotic symmetry generated by $P_{\hat\xi, \hat\xi^i}+G_{\hat{\phi}, \hat{\mu}}$  have the following leading terms in their asymptotic behaviour
\begin{flalign}
	& \hat Y^A  = 0,  \qquad
	\hat b  = 0, \\
	& \hat T  = \frac{i}{8}\delta_{\mathfrak{A} \mathfrak{B}}(\varepsilon_{0,1}^{\mathfrak{A}})^T \varepsilon^{\mathfrak{B}}_{0,2}  - (1 \leftrightarrow 2),\\
	&\hat W  =  \frac{i}{8} \delta_{\mathfrak{A} \mathfrak{B}}(\varepsilon_{0,1}^{\mathfrak{A}})^T  \gamma_0 \gamma_i n^i\varepsilon^{\mathfrak{B}}_{0,2}  - (1 \leftrightarrow 2), \\
	&\hat{\xbar  \phi}  = \frac{i}{8} \left(\varepsilon_{0,1}^{ \mathfrak{B}}\right)^T \gamma_0\varepsilon_{0,2}^{\mathfrak{D}} \epsilon_{\mathfrak{B} \mathfrak{D}} - (1 \leftrightarrow 2), \\
	& \hat \mu = 0
\end{flalign}
Again, only the zero modes of the infinite-dimensional bosonic symmetry group appear: standard space and time translations and global $u(1)$ transformation. 

The term 
\be C_{\varepsilon_1^{\mathfrak{A}}, \varepsilon_1^{\mathfrak{A}}} = \frac{i}{8} \left(\varepsilon_{0,1}^{ \mathfrak{B}}\right)^T \gamma_0 \gamma_5 \varepsilon_{0,2}^{\mathfrak{D}} \epsilon_{\mathfrak{B} \mathfrak{D}} \oint d^2S_i \beta^i
\ee
 is a central charge, which can occur according to general theorems on canonical realizations of asymptotic symmetries \cite{Brown:1986ed,Brown:1986nw}, and which does indeed occur herein the canonical formulation, when there are non zero fluxes (see \cite{Henneaux:1999ct} for a canonical derivation).  
 
In a duality-invariant formulation of the electromagnetic sector \cite{Deser:1976iy,Deser:1997mz}, there would be a second vector potential $Z_i$ with its independent gauge invariance $\delta_{\tilde \phi} Z_i = \partial_i \tilde \phi$, giving a second angle-dependent $u(1)$ symmetry  visible either at null \cite{Strominger:2015bla} or spacelike infinity \cite{1790958}.  Duality invariance appears then as a standard Noether symmetry bringing also its canonical charge-generator in the symmetry algebra.

\section{Conclusions}
\label{sec:Conclusions}

We have given here precise boundary conditions on the dynamical fields of $N=1$ supergravity that provide a realization of the super-BMS algebra of \cite{Awada:1985by}, which is characterized by a finite number of fermionic generators. We have then generalized the analysis to extended supergravity models (specifically, $N=2$).

The boundary conditions on the spinors are the same as the ones given in \cite{Teitelboim:1977hc} and involve no parity condition.   We have completely checked that these are compatible with the twisted parity conditions of gravity (coupled to the Maxwell field) \cite{Henneaux:2018hdj,Henneaux:2019yax,Henneaux:2018gfi}, going thereby beyond the analysis of \cite{Regge:1974zd,Teitelboim:1977hc} where the only bosonic generators were found to be just the Poincar\'e ones (without BMS supertranslations).   This is what allows to get the super-BMS algebra (and not just the super-Poincar\'e ones).

Recent investigations of soft theorems for supergravity suggest that a bigger extension of the BMS algebra, with an infinite number of fermionic generators, should play a physically relevant role \cite{Strominger:2017zoo}.  The  existence of such an intriguing extension was actually already considered in \cite{Awada:1985by} on  algebraic grounds, as a consistent algebraic extension of the BMS algebra.  This raises the question as to whether consistent boundary conditions can be devised that would realize that bigger algebra at spatial infinity.  One promising possibility would be to allow a $O(\frac1r)$ term in the gravitino field $\psi_k$, of the form of a gradient.  This improper gauge term would decay slowlier than the ``core'' term contributing to the charge, as in the case of electromagnetism in higher dimensions \cite{Henneaux:2019yqq}.  It is hoped to return to this question in the future.

\section*{Acknowledgements}
We thank C\'edric Troessart for important discussions.
This work was partially supported by the ERC Advanced Grant ``High-Spin-Grav'' and by FNRS-Belgium (convention IISN 4.4503.15).

\begin{appendix}

\section{Some technical steps in the derivation of  (\ref{eq:PBAlgebraSUSY})} 
\label{sec:appSusyBra}

In this appendix, we illustrate the derivation of the Poisson bracket algebra (\ref{eq:PBAlgebraSUSY}) of Section \ref{sec:AsymptoticAlgebra}
by considering the relations involving the spinor parameters, which are new with respect to the pure gravity case.  Relation (\ref{eq:TransfSusyCharge}) simply expresses that the supercharge is a Lorentz spinor under the compensating local Lorentz transformations that accompany asymptotic spacetime Lorentz transformations, so we focus on (\ref{eq:hamilbmsIII}) and (\ref{eq:hamilbmsIV}), which provide the relevant information enabling the identification with the super-algebra of \cite{Awada:1985by}.

We wish to determine the leading order ($O(1)$) of the anticommutator of two supersymmetry transformations. We will do this by evaluating this anticommutator on the bosonic variables $h_{ij}$, which provides the full information.  There is a subtlety, however, in the derivation: it is that the leading order of $h_{ij}$ is invariant under $O(1)$ diffeomorphisms, so that we must keep the first subleading term in order to read off the asymptotic value of the anticommutator.   More precisely, under $O(1)$ transformations, 
\begin{eqnarray} 
&& \xi = \xi^{(0)} + \xi^{(1)} + O\left(\frac{1}{r^2}\right) \\
&& \xi^i = \xi^{i (0)} + \xi^{i (1)} + O\left(\frac{1}{r^2}\right)
\end{eqnarray}
one has
\begin{eqnarray}
\delta h^{(2)}_{ij} &=& - 2 \xi^{(0)} K_{ij}^{(2)} \nonumber \\
&&+ \xi^{k (0)} \d_k h^{(1)}_{ij} + \d_i  \xi^{k (0)} h^{(1)}_{kj}  + \d_j  \xi^{k (0)} h^{(1)}_{ik} \qquad \nonumber \\
&& + \partial_i \xi^{(1)}_j + \partial_j \xi^{(1)}_i \label{eq:ImproperAndProper}
\end{eqnarray} 
In this appendix,  indices in parentheses on a quantity systematically denote the order in $r^{-1}$.  The corresponding power of $r^{-1}$ is included, so for instance,
$ \xi^{(1)} $ has the form 
$$
\xi^{(1)} = \frac{\tilde \xi ({\mathbf n})}{r}
$$
for some function of the angles $\tilde \xi ({\mathbf n})$ and similarly
$$
K_{ij}^{(2)} = \frac{\xbar K_{ij} ({\mathbf n})} {r^2} , \qquad h^{(1)}_{ij} = \frac{\xbar h_{ij} ({\mathbf n})} {r}
$$
etc.

The terms $\xi^{(1)}$ and $ \xi^{i (1)}$ define proper gauge transformations, which can be present in the transformation of $h^{(2)}_{ij} $ as (\ref{eq:ImproperAndProper}) shows.

On account of the Jacobi identity, one has for a general phase space function $F$,
\begin{eqnarray}
&& \Big\{F, \{ S_{\varepsilon_1}, S_{\varepsilon_2} \} \Big\} \nonumber \\
&& \qquad = \Big\{\{ F,  S_{\varepsilon_1}\}, S_{\varepsilon_2} \Big\} - \Big\{\{ F,  S_{\varepsilon_2}\}, S_{\varepsilon_1} \Big\} \qquad
\end{eqnarray}
The strategy is to compute the left-hand side of this equation from the right-hand side, which is explicitly know for $F = h_{ij}$.  Indeed, $\{ h_{ij},  S_{\varepsilon_1}\} = (i/4)(\xbar \varepsilon_1 \gamma_i \psi_j + \xbar \varepsilon_1 \gamma_j \psi_i )$ and therefore $\{ \{ h_{ij},  S_{\varepsilon_1}\}, S_{\varepsilon_2}\}= (i/4) (\xbar \varepsilon_1 \gamma_i \delta_{\varepsilon_2} \psi_j + \xbar \varepsilon_1 \gamma_j \delta_{\varepsilon_2} \psi_i)$ where we need to keep terms only up to order $O(r^{-2})$.  Using (\ref{eq:TRGravitino}), one finds explicitly three types of terms in $(i/4) \xbar \varepsilon_1 \gamma_i \delta_{\varepsilon_2} \psi_j$, namely
$$
\frac{i}{4} \xbar \varepsilon_1 \gamma_i  \partial_j \varepsilon_2 , \quad - \frac{i}{16} \xbar \varepsilon_1 \gamma_i \d_r h_{js} \gamma^{rs} \varepsilon_2, \quad \frac{i}{8}  \xbar \varepsilon_1 \gamma_i K_{js} \gamma^0 \gamma^s \varepsilon_2.
$$

Since $\d_j \varepsilon_2$ is order $O(r^{-2})$, the first term can be written, to that same order,   $(i/4) \xbar \varepsilon_{0,1} \gamma_i  \partial_j \varepsilon_2^{(1)} = \d_j ((i/4) \xbar \varepsilon_{0,1} \gamma_i  \varepsilon_2^{(1)} )$.  [To the order being considered, we can identify the gamma matrices with coordinate and internal indices, i.e., $\gamma_i$ can be replaced by the constant flat space $\gamma_i$-matrix. We shall thus not make the distinction in this computation.]  When symmetrized over $(i,j)$ this term is a proper gauge transformation and we can therefore neglect it since we want to determine the leading order of the diffeomorphism.

The second term yields, after substraction of the same term coming from $\{\{ h_{ij},  S_{\varepsilon_2}\}, S_{\varepsilon_1}\}$,
\begin{eqnarray}
&& \frac{i}{16}\Big( \xbar \varepsilon_1(\gamma^{rs} \gamma_i - \gamma_i  \gamma^{rs}) \varepsilon_2 \Big)  \d_r h_{js} \nonumber \\
&&  \qquad  \qquad = \frac{i}{8} (\xbar \varepsilon_1\gamma^r \varepsilon_2 \d_r h_{ij} - \xbar \varepsilon_1\gamma^s \varepsilon_2 \d_i h_{sj}). \nonumber
\end{eqnarray}
To the relevant order $O(r^{-2})$, we need to keep only the $O(1)$ terms $\varepsilon_{0,1}$ and $\varepsilon_{0,2}$ of the supersymmetry parameters, which are constant, and the $O(r^{-1})$ term $h^{(1)}_{ij} $ of $h_{ij}$.  Up to irrelevant proper gauge diffeomorphisms, this gives
$$ 
\delta h^{(2)}_{ij}  = \hat W^k \d_k h^{(1)}_{ij}, \quad \hat W^k = \frac{i}{8} \xbar \varepsilon_{0,1}\gamma^k \varepsilon_{0,2} = \frac{i}{8}  \varepsilon_{0,1}^T \gamma_0\gamma^k \varepsilon_{0,2}
$$
which is precisely of the form of the second line of (\ref{eq:ImproperAndProper})  since $\partial_i \hat W^k = 0$. One can write $\hat W_k = \d_k (r \hat W)$ with 
$$r \hat W = \frac{i}{8}  \varepsilon_{0,1}^T \gamma_0\gamma_k x^k\varepsilon_{0,2}, $$ leading to the expression given above.

We turn lastly to the third term, which yields
$$
-\frac{i}{8} \varepsilon_1^T (\gamma_i  \gamma^s + \gamma^s \gamma_i)\varepsilon_2 K_{js}  = - \frac{i}{8}\varepsilon_1^T \varepsilon_2 K_{ij}.
$$
When symmetrized over $(i,j)$, we get to the relevant order a term of the form of the first line of (\ref{eq:ImproperAndProper}), with $$\xi^{(0)} \equiv \hat T = \frac{i}{8} \varepsilon_{0,1}^T \varepsilon_{0,2} .$$

This completes the computation of the Poisson bracket of two supersymmetry generators.

\end{appendix}


\begin{thebibliography}{99}
\bibitem{Conventions}  
 Although not necessary, we impose the ``time gauge'' condition which states that the temporal leg $e^0_\mu$ of the tetrad is orthogonal to the constant time hypersurfaces, $e^0_k = 0$.  The three vectors $e_1^\mu$, $e_2^\mu$, $e_3^\mu$ are then tangent to these hypersurfaces. The local Lorentz gauge freedom is thus reduced to local $SO(3)$ rotations.  Unless otherwise mentioned, objects with spatial indices are three-dimensional objects, e.g., $g^{km}$ is the three-dimensional inverse to the spatial metric $g_{km}$, $g^{km}g_{mr} = \delta^k_r$ (and in general, $g^{km} \not= \! \!^{(4)}g^{km}$).  The triad $e_a^k$ ($a$ internal index) is a orthonormal frame for the spatial metric,
$$e^k_a e_b^m g_{km} = \delta_{ab}.$$  (The quantity $\! ^{(4)} e^k_a$ actually coincides with $e^k_a$ since these vectors are tangent to the equal time hypersufaces.)  Useful relations are
$$ e^k_a e_b^m \delta^{ab} = g^{km} , \; \;  e_k^a e_m^b \delta_{ab} = g_{km} , \; \; e_k^a e_m^b g^{km}  = \delta^{ab} .$$
One also finds for $e^0_\mu$ and $e_0^\mu$ ($0$ = internal index)
$$ e_\mu^0: \; e_k^0 = 0, \; e_0^0 = N; \quad e^\mu_0: \; e_0^0  = \frac{1}{N}, \; e_0^k = - \frac{N^k}{N}.$$
Other useful relations are
$ e^a_0 =  N^k e^a_k$, $e_a^0 = 0$ ($a$ internal index, $0$ world index).


Concerning the spinors, we take a Majorana representation for the Dirac $\gamma$-matrices, which fulfill
$$ \gamma_\Delta \gamma_\Xi + \gamma_\Xi \gamma_\Delta = 2 \eta_{\Delta \Xi} \quad (\Delta, \Xi = 0, 1,2,3)$$
($\Delta=$ internal Lorentz index $=0,a$) where the Minkowski metric has the mostly $+$ signature $(-,+,+,+)$.   The $\gamma$-matrices are thus real, with symmetric $\gamma_1$, $\gamma_2$, $\gamma_3$ and antisymmetric $\gamma_0$.   We define $\gamma_5 = \gamma_0 \gamma_1 \gamma_2 \gamma_3$ and one has $\left(\gamma_5 \right)^2 = - I$, $\left(\gamma_5 \right)^T = - \gamma_5$.  We also define $\gamma_{\Gamma \Delta} = \frac12 \left(\gamma_\Gamma \gamma_\Delta - \gamma_\Delta \gamma_\Gamma \right)$, as well as $\gamma_{\Gamma \Delta \Xi } =  \gamma_{[\Gamma} \gamma_\Delta \gamma_{\Xi]} = \epsilon_{\Gamma \Delta \Xi \Psi} \gamma_5 \gamma^\Psi$ with $\epsilon_{0123} = +1$. For Majorana spinors, the Dirac adjoint becomes $\xbar \psi = \psi^T \gamma_0$.  We convert local indices to ``world'' indices using the tetrads, e.g., $\gamma_\mu = e^\Delta_\mu \gamma_\Delta$, $\gamma_\Delta = e_\Delta^\mu \gamma_\mu$.  One gets $ \gamma_\mu \gamma_\nu + \gamma_\nu \gamma_\mu = 2 g_{\mu \nu}$.
\end{thebibliography}
\end{document}